\documentclass[aps,draft,prb]{revtex4}

\begin{document}
\draft
\title{ Signature of entangled eigenstates in the magnetic response of two coupled flux qubits. }
\author{A.Yu.~Smirnov}
\affiliation{ D-Wave Systems Inc. 320-1985 West Broadway,\\
 Vancouver, B.C. V6J 4Y3, Canada   }

\date{\today }

\begin{abstract}
We study dissipative dynamics and a magnetic response of two coupled flux qubits interacting with a high quality tank in the framework of the impedance measurement technique (IMT). 
It is shown that the observation of the difference between a sum of IMT signals from separated qubits and the signal  from the system when both qubits are in the degeneracy point (IMT deficit) implies immediately a formation of the entangled two-qubit eigenstates. 

\end{abstract}

%\pacs{03.67.Mn, 03.67.Lx, 03.65.Yz, 85.25.Cp}

\maketitle

The impedance measurement technique (IMT) \cite{Greenberg2002,APL_IMT} has proven to be among effective methods for experimental investigation of flux quantum bits. In the framework of this approach the qubit is inductively coupled to the high quality tank (LC circuit) that modifies the impedance of the circuit and, therefore, the angle between the tank voltage and the bias current applied to the tank. This modification is proportional to the second derivative of the qubit's energy or to the magnetic suceptibility of the qubit \cite{ASRabi2003}. The IMT method was succesfully employed for studying the single flux qubits in equilibrium \cite{Landau-Zener,IMT2003} and strongly non-equilibrium states \cite {Rabi}. 
Recently, the IMT measurements were performed for the system of two inductively coupled flux qubits \cite{Izmalkov1} with independent control of the bias in each qubit. The areas of the qubits were practically the same. Because of this, it was necessary to apply an additional flux created by a supplementary wire to change a relative position of the IMT dips on a graph describing a current-voltage dependence as a function of a common bias produced by the dc component of the current in the tank.  Applying the additional flux allows to bring the IMT dips from the different qubits into coincidence (overlapped regime) as well as to pull them apart (regime of separated qubits). 
Effects of interaction and entanglement of the qubits have been observed by comparing IMT signals from the separated qubits with the response of the two-qubit system when both qubits were in the degeneracy point. The IMT dip for the system in the overlapped regime is shown to be smaller than the sum of IMT amplitudes from two separated qubits (IMT deficit) that points to the formation of entangled eigenstates of two coupled qubits. 

Here we present detailed calculations of the magnetic response of the two-qubit system coupled to a dissipative environment. Besides that we elucidate a relation between the IMT deficit and 
an entanglement of pure eigenstates of the system.

We consider two coupled flux qubits interacting with a dissipative environment and with a high-quality LC circuit (tank). The Hamiltonian of a such system has the form $H
= H_0 + H_T + H_{int} + H_{diss}, $ where the Hamiltonian of the two-qubit system is 
\begin{equation}\label{eq_H_0} H_0 =-\Delta_a
\sigma_{x}^{(a)} - \Delta_b \sigma_{x}^{(b)} + \epsilon_a
\sigma_{z}^{(a)} + \epsilon_b \sigma_{z}^{(b)} + J
\sigma_{z}^{(a)} \sigma_{z}^{(b)},
\end{equation}
$H_T$ is the Hamiltonian of the tank with the current $I_T$, the
qubit-tank interaction is given by the term 
\begin{equation}\label{eq_H_int}
H_{int}=-(\lambda_a \sigma_{z}^{(a)} + \lambda_b \sigma_{z}^{(b)}
)I_T,
\end{equation}
and $H_{diss} = - \sigma_z^{(a)} Q_a - \sigma_z^{(b)} Q_b  $ describes the weak coupling of the qubits
to the dissipative baths  with variables $Q_a$ and $Q_b$.
The IMT signal \cite{Izmalkov1,ASRabi2003},
\begin{equation}\label{eq_tan}
\tan\Theta = -2 \frac{Q_T}{L_T} \chi'(\omega_T ),
\end{equation}
 is determined by the real part of the magnetic susceptibility of the double-qubit system, $\chi'(\omega)$, where
\begin{equation} \label{eq_chi}
\chi(\omega) = \int d(t-t_1) e^{i\omega(t-t_1)}\langle \frac{\delta( \lambda_a \sigma_z^{(a)} + 
\lambda_b \sigma_z^{(b)})(t)}{\delta I_T(t_1)}\rangle.
\end{equation}
 In the general case the eigenstate $|\mu \rangle, \mu=1,2,3,4,$ of the Hamiltonian $H_0: H_0|\mu \rangle = E_{\mu}|\mu \rangle,$  can be represented as a superposition of the double-qubit states in the standard basis
\begin{equation}\label{eq_expan} 
|\mu \rangle = c_{1\mu} |00\rangle + c_{2\mu} |01\rangle + c_{3\mu} |10\rangle + c_{4\mu} |11\rangle ,
\end{equation}
where $|00\rangle = |0\rangle_a \otimes |0 \rangle_b$ is a direct product of states of the qubits "a" and "b" with $ \sigma_z^{(a)}|0\rangle_a = - |0\rangle_a,  \sigma_z^{(a)}|1\rangle_a =  |1\rangle_a,$
 and so on. The states $|\mu \rangle $ and $|\nu \rangle$ are orthonormal:
$\langle \mu | \nu \rangle  = \delta_{\mu \nu}.$

We introduce matrix elements for $\sigma_z^{(a)}$ and $\sigma_z^{(b)}$:
\begin{eqnarray}\label{eq_u_v}
 u_{\mu \nu} = \langle \mu |\sigma_z^{(a)}|\nu \rangle = - c_{1\mu}^*c_{1\nu} -  c_{2\mu}^*c_{2\nu} +  c_{3\mu}^*c_{3\nu} + c_{4\mu}^*c_{4\nu}, \nonumber\\
v_{\mu \nu} = \langle \mu |\sigma_z^{(b)}|\nu \rangle = - c_{1\mu}^*c_{1\nu} + c_{2\mu}^*c_{2\nu} -  c_{3\mu}^*c_{3\nu} + c_{4\mu}^*c_{4\nu}.
\end{eqnarray}
With these notations the matrices  $\sigma_z^{(a)}$ and $\sigma_z^{(b)}$ can be represented as $4\times 4$ matrices: 
$ \sigma_z^{(a)} = \sum_{\mu\nu} u_{\mu \nu} \rho^{(\mu \nu)}, \sigma_z^{(b)} = \sum_{\mu\nu} v_{\mu \nu} \rho^{(\mu \nu)}, $ where 
$\rho^{(\mu \nu)} = |\mu \rangle \langle \nu |$ is the matrix with matrix elements 
$\rho^{(\mu \nu)}_{\alpha\beta} = \delta_{\alpha\mu}\delta_{\beta\nu}$. This matrix has 
 only one non-zero (unit) element located on the $\mu \nu$-place. Then, the Hamiltonian H can be represented as 
\begin{equation} \label{eq_H_rho}
H = \sum_{\alpha} E_{\alpha \alpha} \rho^{(\alpha \alpha)} - \sum_{\alpha \beta}(Q_a + \lambda_a I_T)u_{\alpha \beta} \rho^{(\alpha \beta)} +   \sum_{\alpha \beta}(Q_b + \lambda_b I_T)v_{\alpha \beta} \rho^{(\alpha \beta)}.
\end{equation}
The matrices $\rho^{(\mu \nu)}$ can be considered as Heisenberg operators which obey the equations:
\begin{eqnarray}\label{eq_Heis}
i \dot{\rho}^{(\mu \nu)} = - \omega_{\mu \nu} \rho^{(\mu \nu)} - 
\sum_{\alpha}  ( u_{\nu \alpha} Q_a  +  v_{\nu \alpha} Q_b ) \rho^{(\mu \alpha)} + 
\sum_{\alpha}  ( u_{\alpha \mu} Q_a  +  v_{\alpha \mu} Q_b ) \rho^{(\alpha \nu)}  - \nonumber\\
\sum_{\alpha}  (\lambda_a u_{\nu \alpha}  +  \lambda_b v_{\nu \alpha} )I_T \rho^{(\mu \alpha)} + 
\sum_{\alpha}  (\lambda_a u_{\alpha \mu}  +  \lambda_b v_{\alpha \mu} )I_T \rho^{(\alpha \nu)},
\end{eqnarray}
with $\omega_{\mu \nu} = E_{\mu} - E_{\nu}. $ For the response of the heat baths on the action of the qubits we obtain from the theory of open quantum systems \cite{ES1981,ASRabi2003}
\begin{eqnarray}\label{eq_Q}
Q_a(t) = Q_a^{(0)}(t) + \int dt_1 \{ u_{\alpha'\beta'} \varphi_{aa}(t,t_1) + v_{\alpha'\beta'} \varphi_{ab}(t,t_1) \} \rho^{(\alpha'\beta')}(t_1) \}, \nonumber\\
Q_b(t) = Q_b^{(0)}(t) + \int dt_1 \{ u_{\alpha'\beta'} \varphi_{ba}(t,t_1) + v_{\alpha'\beta'} \varphi_{bb}(t,t_1) \} \rho^{(\alpha'\beta')}(t_1) \}.
\end{eqnarray}
Here $Q_a^{(0)}, Q_b^{(0)} $ are unperturbed variables of the heat baths, 
\begin{equation}\label{eq_varphi_aa}
\varphi_{aa}(t,t_1) = \langle i [ Q_a^{(0)}(t), Q_a^{(0)}(t_1)]_- \rangle \theta(t-t_1) = 
\int \frac{d\omega}{2\pi} e^{-i\omega (t-t_1)} \chi_{aa}(\omega)
\end{equation}
is the linear response function and the susceptibility of the "a" heat bath,  
\begin{equation}\label{eq_varphi_ab}
\varphi_{ab}(t,t_1) = \langle i [ Q_a^{(0)}(t), Q_b^{(0)}(t_1)]_- \rangle \theta(t-t_1)
\end{equation}
is a response function describing a cross-correlation between "a" and "b" heat baths with the corresponding susceptibility $\chi_{ab}(\omega)$,  $\theta(\tau ) $ is the Heaviside step function.  
We introduce also corresponding correlation functions and spectra 
\begin{eqnarray}\label{eq_M}
M_{aa}(t,t_1) = \langle (1/2) [ Q_a^{(0)}(t), Q_a^{(0)}(t_1)]_+ \rangle = \int \frac{d\omega}{2\pi} e^{-i\omega (t-t_1)} S_{aa}(\omega)
, \nonumber\\
M_{ab}(t,t_1) = \langle (1/2) [ Q_a^{(0)}(t), Q_b^{(0)}(t_1)]_+ \rangle = \int \frac{d\omega}{2\pi} e^{-i\omega (t-t_1)} S_{ab}(\omega)
\end{eqnarray}
together with combinations  
\begin{eqnarray}\label{eq_combin}
\varphi^{\alpha'\beta'}_{\nu \alpha}(t,t_1) = 
\langle i [ u_{\nu \alpha} Q_a^{(0)}(t) + v_{\nu \alpha} Q_b^{(0)}(t), 
u_{\alpha' \beta'} Q_a^{(0)}(t_1) + v_{\alpha'\beta'} Q_b^{(0)}(t_1) ] _-\rangle  \theta(t-t_1), 
\nonumber\\
M^{\alpha'\beta'}_{\nu \alpha}(t,t_1) = 
\langle (1/2) [ u_{\nu \alpha} Q_a^{(0)}(t) + v_{\nu \alpha} Q_b^{(0)}(t), 
u_{\alpha' \beta'} Q_a^{(0)}(t_1) + v_{\alpha'\beta'} Q_b^{(0)}(t_1) ] _+\rangle.
\end{eqnarray} 
With these definitions the "collision" terms in Eq.~(\ref{eq_Heis}), averaged over the thermodynamically equilibrium state of the heat baths, take the form
\begin{eqnarray}\label{eq_collis}
 \langle (1/2)[u_{\nu \alpha} Q_a(t) + v_{\nu \alpha} Q_b(t), \rho^{(\mu \alpha)}(t)]_+ \rangle = \nonumber\\
\int dt_1  \tilde{M}^{\alpha'\beta'}_{\nu \alpha}(t,t_1) \langle i [ \rho^{(\mu\alpha)}(t), \rho^{(\alpha'\beta')}(t_1)]_-\rangle +  
\int dt_1  \varphi^{\alpha'\beta'}_{\nu \alpha}(t,t_1) \langle (1/2) [ \rho^{(\mu\alpha)}(t), \rho^{(\alpha'\beta')}(t_1)]_+\rangle ; \nonumber\\
\langle (1/2)[u_{\alpha\mu} Q_a(t) + v_{\alpha\mu} Q_b(t), \rho^{\alpha\nu}(t)]_+ \rangle = \nonumber\\
\int dt_1  \tilde{M}^{\alpha'\beta'}_{\alpha \mu}(t,t_1) \langle i [ \rho^{(\alpha \nu)}(t), \rho^{(\alpha'\beta')}(t_1)]_-\rangle +  
\int dt_1  \varphi^{\alpha'\beta'}_{\alpha \mu}(t,t_1) \langle (1/2) [ \rho^{(\alpha \nu)}(t), \rho^{(\alpha'\beta')}(t_1)]_+\rangle ,
\end{eqnarray}
where $\tilde{M}(\tau) = M(\tau) \theta(\tau).$
In the case of weak interactions between the qubits and the baths we can use a free evolution of the matrices $\rho^{(\mu \alpha)}(t) = e^{i\omega_{\mu\alpha}(t-t_1)}\rho^{(\mu \alpha)}(t_1) $ and  
$ \rho^{(\alpha \nu)}(t) = e^{i\omega_{\alpha\nu}(t-t_1)} \rho^{(\alpha \nu)}(t_1)  $  to calculate 
(anti)commutators in the collision terms (\ref{eq_collis}) using the relations: 
$$ [\rho^{(\mu \alpha)}(t_1), \rho^{(\alpha'\beta')}(t_1)]_{\pm} = \delta_{\alpha'\alpha}\rho^{(\mu \beta')}(t_1) \pm \delta_{\beta'\mu} \rho^{(\alpha'\alpha)}(t_1)$$
$$
 [\rho^{(\alpha \nu)}(t_1), \rho^{(\alpha'\beta')}(t_1)]_{\pm} = \delta_{\alpha'\nu}\rho^{(\alpha \beta')}(t_1) \pm \delta_{\beta'\alpha} \rho^{(\alpha'\nu)}(t_1), $$ where the all Heisenberg matrices are taken at the same moment of time $t_1.$ As a result we obtain the following equation for the averaged operators $\langle \rho^{(\mu \nu)} \rangle $ (we consider thereafter the evolution of the averaged variables only, and, because of this, omit the brackets $\langle ..\rangle $):
\begin{eqnarray}\label{eq_av_rho}
\dot{\rho}^{(\mu\nu)} = i \omega_{\mu\nu}\rho^{(\mu\nu)} + i \sum_{\alpha}\left[(\lambda_a u_{\nu\alpha}+ \lambda_b v_{\nu\alpha}) \rho^{(\mu\alpha)} - (\lambda_a u_{\alpha\mu}+ \lambda_b v_{\alpha\mu}) \rho^{(\alpha\nu)}\right]I_T - \nonumber\\
\sum_{\alpha\beta}\int dt_1 \left( \tilde{M}-\frac{i}{2} \varphi \right)^{\alpha \beta}_{\nu \alpha}(t-t_1) e^{i\omega_{\mu\alpha}(t-t_1)}\rho^{(\mu\beta)}(t_1) + \nonumber\\
\sum_{\alpha\beta} \int dt_1 \left( \tilde{M}+\frac{i}{2} \varphi \right)^{\beta\mu}_{\nu \alpha}(t-t_1) e^{i\omega_{\mu\alpha}(t-t_1)}\rho^{(\beta\alpha)}(t_1) + \nonumber\\
\sum_{\alpha\beta} \int dt_1 \left( \tilde{M}-\frac{i}{2} \varphi \right)^{\nu\beta}_{\alpha\mu}(t-t_1) e^{i\omega_{\alpha\nu}(t-t_1)}\rho^{(\alpha\beta)}(t_1) - \nonumber\\
\sum_{\alpha\beta} \int dt_1 \left( \tilde{M}+\frac{i}{2} \varphi \right)^{\beta\alpha}_{\alpha\mu}(t-t_1) e^{i\omega_{\alpha\nu}(t-t_1)}\rho^{(\beta\nu)}(t_1), 
\end{eqnarray}
with $(\tilde{M}-\frac{i}{2} \varphi )^{\alpha \beta}_{\nu \alpha} = \tilde{M}^{\alpha \beta}_{\nu \alpha} -\frac{i}{2} \varphi^{\alpha \beta}_{\nu \alpha}. $ 
In the equilibrium state, when $I_T=0,$ the averaged matrices $\rho^{(\mu\nu)} = \rho^{(\mu\nu)}_0 $ 
with $\mu\neq \nu$ are equal to zero, whereas for the diagonal matrices $\rho^{(\nu\nu)}_0$ we obtain the equation
%131
\begin{eqnarray}\label{eq_rho_equ1}
\sum_{\alpha} \left[ \left( \tilde{S}+\frac{i}{2}\chi \right)^{\alpha \nu}_{\nu\alpha}(\omega_{\nu\alpha}) + \left( \tilde{S}-\frac{i}{2}\chi \right)^{\nu\alpha}_{\alpha\nu}(\omega_{\alpha\nu})\right]\rho^{(\alpha\alpha)}_0 = \nonumber\\
\rho^{(\nu\nu)}_0  \sum_{\alpha} \left[ \left( \tilde{S}-\frac{i}{2}\chi \right)^{\alpha \nu}_{\nu\alpha}(\omega_{\nu\alpha}) + \left( \tilde{S}+\frac{i}{2}\chi \right)^{\nu\alpha}_{\alpha\nu}(\omega_{\alpha\nu})\right].
\end{eqnarray}
For the heat baths with the same parity regarding the time inversion we have: $\chi_{ab}(\omega) = \chi_{ba}(\omega), S_{ab}(\omega) = S_{ba}(\omega), $ where the spectrum $S_{ab}(\omega)$ is related to the imaginary part of the corresponding susceptibility $\chi_{ab}''(\omega)$ according to the fluctuation-dissipation theorem: $S_{ab} = \chi_{ab}''(\omega) \coth(\omega/2T).$ Here $T$ is the temperature of the heat baths "a" and "b". The same relations take place for $S_{aa}(\omega)$ and $\chi_{aa}''(\omega),$ as well as for $S_{bb}(\omega)$ and $\chi_{bb}''(\omega).$ 
In view of the fact that $u_{\alpha\nu}= u^*_{\nu\alpha}, v_{\alpha\nu} = v^*_{\nu\alpha}$ and with the notation $\rho^{(\nu\nu)}_0\equiv  \rho_{\nu}$ the equation 
(\ref{eq_rho_equ1}) can be rewritten as 
\begin{eqnarray}\label{eq_rho_equ2}
\sum_{\alpha}\left\{ 
\rho_{\alpha} \left[\coth\left(\frac{\omega_{\nu\alpha}}{2T}\right)-1\right] - 
\rho_{\nu} \left[\coth\left(\frac{\omega_{\nu\alpha}}{2T}\right)+1\right] \right\} \times 
\nonumber\\
 \{ |u_{\nu \alpha}|^2\chi''_{aa}(\omega_{\nu\alpha}) + |  v_{\nu \alpha}|^2\chi''_{bb}(\omega_{\nu\alpha}) + 
(u_{\nu \alpha}v^*_{\nu\alpha} + v_{\nu\alpha}u_{\nu\alpha}^*)\chi''_{ab}(\omega_{\nu\alpha}) \} = 0.
\end{eqnarray}
The solution of this equation is given by the Gibbs distribution: $ \rho_{\nu} = \exp(-E_{\nu}/T)/Z,$ where $Z=\sum_{\alpha}\exp(-E_{\alpha}/T).$ 
The term, related to the interaction of qubits with the tank's current $I_T$ in Eq.(\ref{eq_av_rho}),
transforms to the form: $ i(\lambda_a u_{\nu \mu} + \lambda_b v_{\nu\mu})(\rho_{\mu}-\rho_{\nu})I_T.$ The interaction with the tank drops out from the equation for diagonal elements $\rho^{(\nu\nu)}$, so that $\langle \delta\rho^{(\nu\nu)}(t)/\delta I_T(t')\rangle = 0.$ 
It means that only non-diagonal elements, $\rho^{(\mu\nu)},$ give non-zero contributions to the susceptibility $\chi(\omega)$ (\ref{eq_chi}). For the Fourier transform of the functional derivative 
$\langle \delta \rho^{(\mu\nu)}(t)/\delta I_T(t')\rangle$ we obtain from Eq. (\ref{eq_av_rho}):
\begin{equation}\label{eq_rhoFD}
\langle \frac{\delta\rho^{(\mu\nu)}}{\delta I_T}(\omega) \rangle = i \frac{\lambda_a u_{\nu\mu}+\lambda_b v_{\nu\mu} }{\omega + E_{\mu} - E_{\nu}  + i\Gamma_{\mu\nu}(\omega) } (\rho_{\nu} - \rho_{\mu} ). 
\end{equation}
A small frequency shift of the double-qubit system caused by the dissipative environment is omitted here. With a notation
\begin{equation}\label{eq_Lam}
\Lambda_{\nu\alpha}(\omega ) = |u_{\nu \alpha}|^2\chi''_{aa}(\omega) + 
|v_{\nu \alpha}|^2\chi''_{bb}(\omega) + 
(u_{\nu \alpha}v^*_{\nu\alpha} + v_{\nu\alpha}u_{\nu\alpha}^*)\chi''_{ab}(\omega) 
\end{equation}
the frequency-dependent decoherence rate is given by the expression
\begin{eqnarray}\label{eq_Gamma1}
\Gamma_{\mu\nu}(\omega) = \sum_{\alpha}\frac{1}{2} \left[\coth\left(\frac{\omega + \omega_{\mu\alpha}}{2T}\right)+1\right] \Lambda_{\nu\alpha}(\omega + \omega_{\mu\alpha} ) + \nonumber\\
\sum_{\alpha}\frac{1}{2} \left[\coth\left(\frac{\omega + \omega_{\alpha\nu}}{2T}\right)-1\right] \Lambda_{\mu\alpha}(\omega + \omega_{\alpha\nu} ). 
\end{eqnarray}
Because of the frequency dependence of $ \Gamma_{\mu\nu}(\omega) $ the dissipative evolution of the averaged elements, $\langle \rho^{(\mu\nu)}(t) \rangle ,$ is governed by the decoherence rate, $ \Gamma_{\mu\nu}(E_{\nu} - E_{\mu}),$ which differs from the decoherence rate $\Gamma_{\mu\nu}(\omega_T)$ incorporatED into the low-frequency susceptibility $\chi(\omega_T)$ (\ref{eq_chi}) of the coupled qubits. Taking into account the relations  $ \sigma_z^{(a)} = \sum_{\mu\nu}u_{\mu \nu} \rho^{(\mu \nu)}, \sigma_z^{(b)} = \sum_{\mu\nu} v_{\mu \nu} \rho^{(\mu \nu)}, $ for the magnetic susceptibility we find the equation
\begin{eqnarray}\label{eq_chiF}
\chi(\omega) = \sum_{\mu\neq\nu} (\lambda_a u_{\mu\nu}+ \lambda_b v_{\mu\nu}) \langle \frac{\delta\rho^{(\mu\nu)}}{\delta I_T}(\omega) \rangle = \nonumber\\ \sum_{\mu\neq\nu}
\frac{\rho_{\nu} - \rho_{\mu} }{ \omega + E_{\mu} - E_{\nu}  + i\Gamma_{\mu\nu}(\omega) } (\lambda_a u_{\mu\nu}+ \lambda_b v_{\mu\nu}) 
(\lambda_a u_{\nu\mu}+\lambda_b v_{\nu\mu}). 
\end{eqnarray}
If we recall that $u_{\mu\nu}$ and $v_{\mu\nu} $ are the matrix elements of the $\sigma_z^{(a)}$ and 
$\sigma_z^{(b)}$ (\ref{eq_u_v}) between the eigenstates $|\mu\rangle$ and $|\nu\rangle$ of the Hamiltonian $H_0$ (\ref{eq_H_0}) and neglect the resonant frequency of the tank compared to the energy differences between the qubit's levels, $\omega_T \ll E_{\nu} - E_{\mu},$ then the current-voltage angle in the tank (\ref{eq_tan}) is given by the expression
\begin{equation}\label{eq_tanF}
\tan \Theta = - 4 \frac{Q_T}{L_T}\sum_{\mu < \nu} \frac{ \rho_{\mu} - \rho_{\nu} }{ E_{\nu} - E_{\mu} } R_{\mu\nu},
\end{equation}
where  
\begin{eqnarray}\label{eq_R}
R_{\mu \nu} = \lambda_a^2 \langle \mu | \sigma_{z}^{(a)}|\nu \rangle \langle \nu |\sigma_{z}^{(a)}|\mu \rangle + \lambda_b^2 \langle \mu |\sigma_{z}^{(b)}|\nu \rangle \langle \nu |\sigma_{z}^{(b)}|\mu \rangle  \nonumber\\
+ \lambda_a \lambda_b \langle \mu |\sigma_{z}^{(a)}|\nu \rangle
\langle \nu |\sigma_{z}^{(b)}|\mu \rangle + 
\lambda_a \lambda_b
\langle \mu |\sigma_{z}^{(b)}|\nu \rangle \langle \nu
|\sigma_{z}^{(a)}|\mu \rangle.
\end{eqnarray}
For the weak qubit-bath couplings the decoherence rates, $\Gamma_{\mu\nu}(\omega_T) \ll E_{\nu} - E_{\mu},$ drop out of Eq.(\ref{eq_tanF}), which describes the IMT signal.

The first line in Eq. (\ref{eq_R}) corresponds to the contribution of individual qubits to the IMT signal. It is clear that this contribution survives even when the eigenstates $|\mu\rangle, |\nu \rangle$ are products of single-qubits states. The terms in the second line of Eq. (\ref{eq_R}) are related to the simultaneous flipping of both qubits.
This term vanishes when all eigenstates of the two-qubit system are non-entangled.
To show that we suppose that the states $|\mu\rangle, |\nu \rangle $ can be represented as products of the wave fuctions related to the qubits "a" and "b": $ |\mu \rangle =|\mu \rangle_a |\mu \rangle_b,  |\nu \rangle =|\nu \rangle_a |\nu \rangle_b.$ It is the case, for example, for the separated qubits, when one qubit is in the degeneracy point, but another - far out of degeneracy. 
The states $|\mu\rangle$ and $|\nu \rangle$ are orthonormal, so that for $\mu\neq \nu$ we have $ \langle \mu |\nu\rangle = 
\langle \mu |\nu\rangle_a \langle \mu |\nu\rangle_b =  0. $  It means that  $\langle \mu |\nu\rangle_a =0$ or $\langle \mu |\nu\rangle_b = 0$, or both. 
Then the matrix elements 
$\langle \mu |\sigma_{z}^{(a)}|\nu \rangle, \langle \nu |\sigma_{z}^{(b)}|\mu \rangle $  
 involved in the expression for 
the IMT deficit \cite{Izmalkov1}, 
\begin{equation}\label{eq_deficit}
 (R_{\mu \nu})_{def} = \lambda_a \lambda_b \{ \langle \mu |\sigma_{z}^{(a)}|\nu \rangle \langle \nu |\sigma_{z}^{(b)}|\mu \rangle + 
\langle \mu |\sigma_{z}^{(b)}|\nu \rangle \langle \nu
|\sigma_{z}^{(a)}|\mu \rangle\},
\end{equation}
have the form: $ \langle \mu |\sigma_{z}^{(a)}|\nu \rangle = \langle \mu |\sigma_{z}^{(a)}|\nu \rangle_a \langle \mu |\nu \rangle_b $ and $ \langle \nu |\sigma_{z}^{(b)}|\mu \rangle = 
\langle \nu |\mu \rangle_a \langle \nu |\sigma_{z}^{(b)}|\mu \rangle_b. $  The products of these matrix elements contain the term $\langle \mu |\nu \rangle_b \langle \nu |\mu \rangle_a $ which should be equal to zero due to the normalization condition, $ \langle \mu |\nu\rangle_a \langle \mu |\nu\rangle_b =  0. $ We can conclude, therefore, that non-entangled eignestates of the two-qubit system gives no contribution to the IMT deficit  $ (R_{\mu \nu})_{def} . $
Thus, the observation of non-zero IMT deficit directly points to a formation of entangled eigenstates in the system of two coupled flux qubits. It should be noted, 
however, that zero IMT deficit does not mean that all eigenstates of the system are non-entangled.

In summary we have performed calculations of the magnetic susceptibility and IMT signals for two coupled flux qubits. We have shown that the observation of non-zero IMT deficit presents a sufficient condition for the existence of entangled eigenstates in the two-qubit system.

Discussions with E. Il'ichev, A. Izmalkov, M. Grajcar, M. Amin, A. Maassen van den Brink, and A. Zagoskin are gratefully acknowledged. I especially thank Alexandre Zagoskin for critical reading of the manuscript.

\end{document}